\documentclass{article}

\usepackage{PRIMEarxiv}

\usepackage[utf8]{inputenc} 
\usepackage[T1]{fontenc} 
\usepackage{url} 
\usepackage{booktabs} 
\usepackage{amsfonts} 
\usepackage{nicefrac} 
\usepackage{microtype} 
\usepackage{lipsum}
\usepackage{fancyhdr} 
\usepackage{graphicx} 
\usepackage{lmodern,textcomp}
\graphicspath{{media/}} 
\usepackage{printlen}

\title{Texture Recognition Using a Biologically Plausible Spiking Phase-Locked Loop Model for Spike Train Frequency Decomposition
}

\author{
 Michele Mastella, Tesse Tiemens, Elisabetta Chicca \\
 BICS, Zernike Inst Adv Mat; CogniGron \\
 University of Groningen  \\
 Groningen, Netherlands\\
 \texttt{\{m.mastella, e.chicca\}@rug.nl; t.e.tiemens@student.rug.nl} \\
}

\usepackage{graphicx}
\graphicspath{{figures/draft/}}

\usepackage{upgreek}
\usepackage{siunitx}
\usepackage{lipsum}
\usepackage[acronym]{glossaries}
\usepackage{amsmath,amsfonts,amssymb}
\usepackage{color}

\usepackage[
backend=biber,
style=nature,
sorting=none
]{biblatex}

\AtEveryBibitem{\clearlist{language}}
\AtEveryBibitem{\clearfield{url}}
\AtEveryBibitem{\clearfield{urlyear}}
\AtEveryBibitem{\clearlist{note}}

\usepackage[acronym]{glossaries}
\newcommand*\myglsentry[1]{%
  \protect\ifglsused{#1}{%
    \glsentryshort{#1}%
  }{%
    \glsentrylong{#1}%
  }%
}

\usepackage[hidelinks]{hyperref}

\newacronym{dpi}{DPI}{Differential Pair Integrator}
\newacronym{asic}{ASIC}{Application Specific Integrated Circuit}
\newacronym{ai}{AI}{Artificial Intelligence}
\newacronym{pmos}{PMOS}{p-type \\myglsentry{cmos} Field Effect Transistor}
\newacronym{nmos}{NMOS}{n-type \\myglsentry{cmos} Field Effect Transistor}
\newacronym{dac}{DAC}{Digital Analog Converter}
\newacronym{tde}{TDE}{Time Difference Encoder}
\newacronym{cmos}{CMOS}{Complementary Metal Oxide Semiconductor}
\newacronym{isi}{ISI}{Interspike Interval}
\newacronym{snn}{SNN}{Spiking Neural Network}
\newacronym{fft}{FFT}{Fast Fourier Transform}
\newacronym{ir}{IR}{Instantaneous Rate}
\newacronym{irh}{IRH}{Instantaneous Rate Histogram}
\newacronym{pll}{PLL}{Phase-Locked Loop}
\newacronym{spll}{sPLL}{Spiking \myglsentry{pll}}
\newacronym{cco}{CCO}{Current Controlled Oscillator}
\newacronym{lif}{LIF}{Leaky Integrate and Fire}
\newacronym{r-lif}{R-LIF}{Recurrent \myglsentry{lif}}
\newacronym{lstm}{LSTM}{Long Short-Term Memory}
\newacronym{fac}{FAC}{Facilitatory Trace}
\newacronym{trg}{TRG}{Trigger Trace}
\newacronym{syn}{SYN}{Synaptic Trace}
\newacronym{ft}{FT}{Fourier transform}
\newacronym{cuba}{CUBA}{Current Based}
\newacronym{bptt}{BPTT}{Backpropagation Through Time}
\newacronym{rnn}{RNN}{Recurrent Neural Network}
\newacronym{nni}{NNI}{Neural Network Intelligence}
\newacronym{sota}{SOTA}{State-Of-The-Art}
\newacronym{dsnn}{D-SNN}{Deep Spiking Neural Network}
\newacronym{sa}{SA}{Slow Adapting Afferents}
\newacronym{fa}{FA}{Fast Adapting Afferent}
\newacronym{pc}{PC}{Pacini Afferent}
\newacronym{dvs}{DVS}{Dynamic Vision Sensor}
\newacronym{srm}{SRM}{Simple Response Model}
\newacronym{nas}{NAS}{Neuromorphic Auditory Sensor}
\newacronym{sfd}{SFD}{Spike Frequency Divider}
\newacronym{slpf}{S-LPF}{Spike-based Low Pass Filter}
\newacronym{shf}{H\&F}{Hold \& Fire}
\newacronym{sig}{I\&G}{Integrate \& Generate}
\newacronym{rbssg}{RBSSG}{Reverse Bitwise Synthetic Spike Generator}
\newacronym{fsm}{FSM}{Finite State Machine}
\newacronym{fpga}{FPGA}{Field Programmable Gate Array}
\newacronym{rf}{R\&F}{Resonate \& Fire}
\newacronym{ttfs}{TTFS}{Time to First Spike}
\newacronym{dft}{DFT}{Digital Fourier Transform}
\newacronym{wta}{WTA}{Winner-Takes-All}
\newacronym{mst}{MST}{Multifrequency Spike Train}
\usepackage{appendix}

\makeglossaries

\begin{document}
\maketitle

\begin{abstract}
In this paper, we present a novel spiking neural network model designed to perform frequency decomposition of spike trains. Our model emulates neural microcircuits theorized in the somatosensory cortex, rendering it a biologically plausible candidate for decoding the spike trains observed in tactile peripheral nerves. We demonstrate the capacity of simple neurons and synapses to replicate the phase-locked loop (PLL) and explore the emergent properties when considering multiple spiking phase-locked loops (sPLLs) with diverse oscillations.

We illustrate how these sPLLs can decode textures using the spectral features elicited in peripheral nerves. Leveraging our model's frequency decomposition abilities, we improve state-of-the-art performances on a Multifrequency Spike Train (MST)  dataset.

This work offers valuable insights into neural processing and presents a practical framework for enhancing artificial neural network capabilities in complex pattern recognition tasks.
\end{abstract}

\twocolumn
\section{Introduction}

Touch is a fundamental sense for most animals. In humans, tactile stimuli are transduced from pressure into electrical signals by four different receptors, known as mechanoreceptors. 
Existing literature establishes a consensus identifying Merkel cells as responsible for encoding constant pressure, Meissner corpuscles as sensitive to pressure variation, Pacini corpuscles as responsive to skin vibrations, and Ruffini corpuscles as related to sustained skin stretching~\cite{Muniak_Ray_ea2007}. 

Merkel cells are typically located on the top layer of the skin, in close proximity to nerve endings. 
The afferents connected to Merkel cells are typically defined as \gls{sa}. 
Models of Merkel cells suggest that they are responsible for converting analog pressure signals into sustained spiking activity, akin to tonic spiking, with the frequency related to the pressure intensity~\cite{Muniak_Ray_ea2007}. 

Meissner cells, predominantly found in glabrous skin, have afferents typically labeled as \glspl{fa}. 
Their ability to detect variations in pressure stimuli, coupled with the highest density of innervation in the fingertip~\cite{Corniani_Saal2020a} (up to 3,000-5,000 Meissner cells per $1\ \mathrm{cm^2}$), makes them well-suited for motion detection and grip control~\cite{Saal_Bensmaia2014a}. 
The activity of their afferents can be linked to the first derivative of the pressure signal.

Pacini corpuscles are located deep within the skin, closer to the bone, and are sensitive to skin vibration, particularly at high frequencies, with the optimum being around \SI{250}{\hertz}~\cite{Saal_Bensmaia2014a}. 
Their response resembles a second derivative of the pressure signal. They can convey information about vibration through a specific afferent usually called \glspl{pc} or \gls{fa}-II~\cite{Muniak_Ray_ea2007}. 

On the other hand, Ruffini corpuscles exhibit an encoding similar to Merkel cells, but instead of pressure, they mainly respond to stretch over the skin. 
Their corresponding afferent is typically called \gls{sa}-II. 
They reside deep in the skin and transfer information about joint movement, likely informing about propioception~\cite{Muniak_Ray_ea2007}. 

The highest density of sustained-responding mechanoreceptors can be found in the fingertip, with approximately \num{10,000} Merkel cells/\unit{\square\centi\meter}, resulting in a average distance greater than \SI{100}{\micro\meter}. In~\cite{Weber_Saal_ea2013}, natural textures are analyzed using a Laser Microscope, highlighting variations on the order of \numrange{10}{100} \unit{\micro\meter} for textures like Nylon or Chiffon. Detection of such fine variations cannot be achieved through simple palpation of textures.

A growing body of evidence in literature is proposing sliding as the primary modality to analyze fine textures~\cite{Weber_Saal_ea2013}. In this scenario, fine textures would elicit complex vibration patterns in the skin. These vibrations would then be detected by \gls{pc} afferents, which could subsequently project to the somatosensory pathway, informing about the texture. This hypothesis was first tested in~\cite{Weber_Saal_ea2013}, where a Macaque had its finger slide on different textures while peripheral nerves connected to either \gls{sa} or \gls{pc} were probed. The spiking activity of these nerves was then analyzed using both spike counts and a \gls{ft}. In the same work, a mean correlation study highlighted that for fine textures, \glspl{pc} analyzed with \gls{ft} were more informative in comparison to \glspl{sa} analyzed with spike count. This would suggest that fine textures are identified by the brain using spectral information carried in single spike trains from the peripheral afferents. 

\begin{figure*}[tbh]
 \centering
 \includegraphics[width=\textwidth]{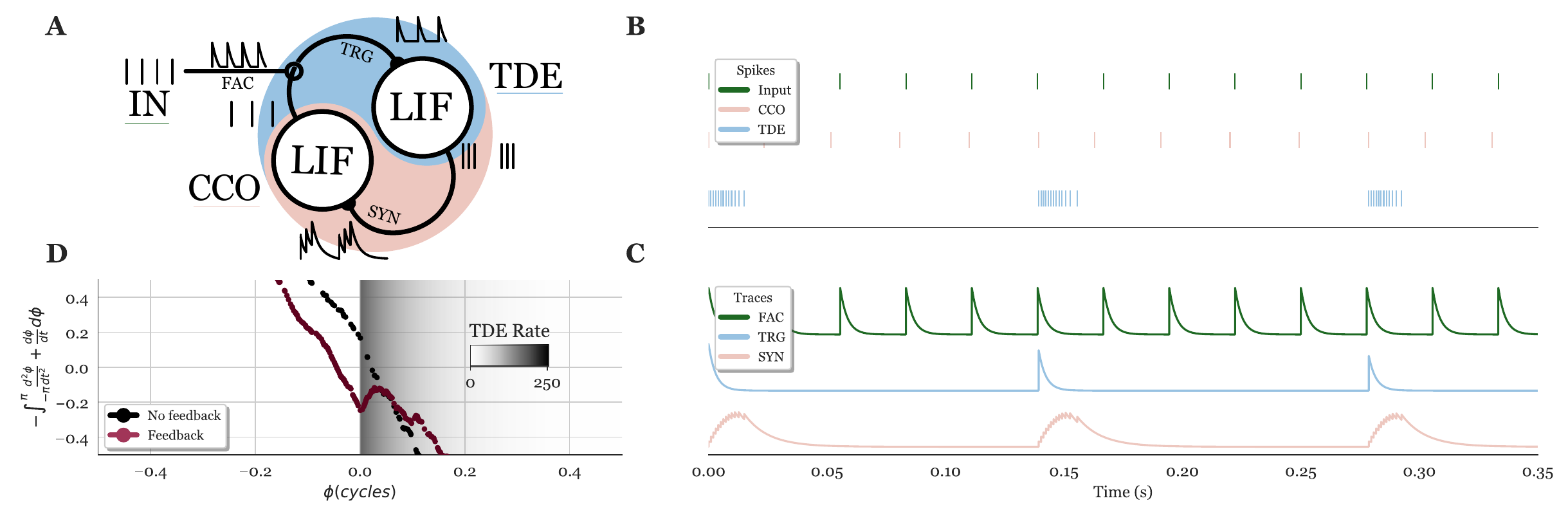}
  \includegraphics[width=\textwidth]{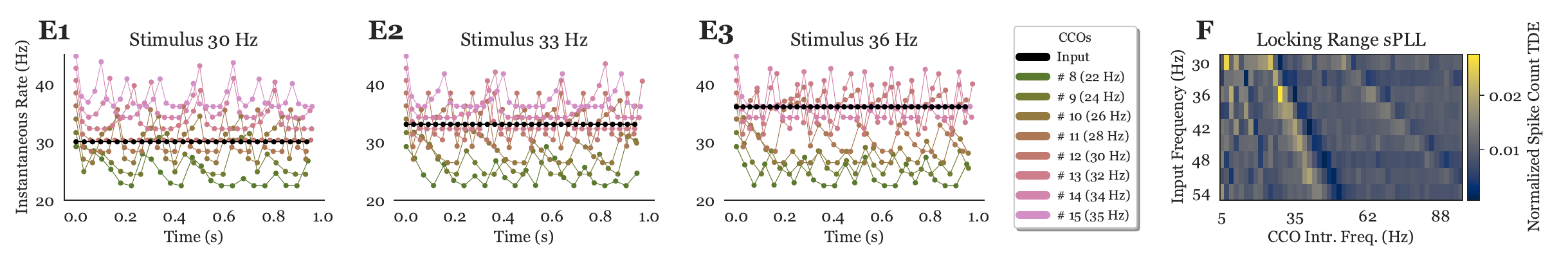}
  \label{fig:spll}
   \caption{In scheme \textbf{A} the \gls{spll} model explained in Sec.~\ref{subsec:spll}. In panel \textbf{B} the spiking activity of the input, the \gls{cco} and the \gls{tde} are depicted for a specific input at \SI{36}{\hertz}. Panel \textbf{C} plots the three synaptic traces in the \gls{spll}. Scatter plot \textbf{D} shows the potential well model explained in Sec.~\ref{subsec:potwell} applied to this specific case, where the simulation was ran for \SI{1}{\second}. In dark red is the case where the feedback is present. In black is the case where the feedback has been removed. The background indicates the \gls{tde} rate intensity. In Figure \textbf{E1, E2, E3} are plotted the response of 13 different \gls{spll} to a spike train with three different frequencies. In the panel \textbf{F}, the normalized spike count for 50 different \glspl{cco} is compared with respect to 27 different input frequencies (\SI{30}{\hertz} to \SI{54}{\hertz}). On the x-axis, it's annotated the frequency at which the different \glspl{cco} spike when no stimulus is presented (called CCO Intrinsic Frequencies). The normalization of the spike count is done over those intrinsic frequencies.}
\end{figure*}

Numerous investigations regarding how the brain decodes textures have been conducted over the years. Accumulating evidence from in-vivo studies on Macaques indicates that to some extent, the brain preserves the precise temporal structure of the spike trains generated in the peripheral afferents~\cite{Weber_Saal_ea2013,Lieber_Bensmaia2019a,Long_Greenspon_ea2022}. This preserved information is then utilized to decode complex textures.

Currently, the mechanism for decoding such stimuli is not fully understood~\cite{Lieber_Bensmaia2019a}. One model, proposed in~\cite{Ahissar1998}, utilizes precise spike times and their associated frequencies to decode textures. In this model, a closed-loop micro-circuit between the cortex and the thalamus is theorized. This micro-circuit implements a phase-locked loop, where the cortex houses controlled oscillators, and the thalamus acts as a phase detector. Neurons in the cortex represent a specific expected frequency (dictated by the anticipated texture), while the thalamus calculates the phase difference between the actual stimulus and the expectation. This algorithm leverages the precise information from the spike train to decode the presented texture.

In the neuromorphic engineering field, several researchers created algorithms for enabling the decoding of textures by artificial agents.
Researchers in John Hopkins University use a dataset created by sliding a robotic finger on several coarse textures~\cite{Nguyen_Osborn_ea2018}. 
The dataset, composed of analog signals recorded from 9 different sensors, is converted to spikes using the Izhikevich neuron model~\cite{Izhikevich2003}. 
A variety of algorithms was applied to the dataset: extreme learning machines~\cite{Rasouli_Chen_ea2018}, sparse representation classification~\cite{Nguyen_Osborn_ea2018}, unsupervised clustering~\cite{Iskarous_Nguyen_ea2018} and support vector machines ~\cite{Sankar_Balamurugan_ea2021}. The textures chosen in these works are coarse and therefore more suited for the approach related to the spatial distribution of the sensors on the artificial skin.

\cite{Gao_Taunyazov_ea2020,Taunyazov_Chua_ea2020} study instead the decoding of fine textures in artificial agents with a custom dataset (BioTac dataset~\cite{Gao_Taunyazov_ea2020}). 
While the first work~\cite{Gao_Taunyazov_ea2020} uses a supervised autoencoder approach, the second~\cite{Taunyazov_Chua_ea2020} converts the analog data recorded by the sensors into spikes using a k-threshold \gls{srm} neuron. The spikes are fed to a \gls{dsnn} composed of \gls{lif} neurons and trained with backpropagation to decode the textures.

Inspired by the work in~\cite{Ahissar1998}, we created a novel model based on the phase-locked loop, which exhibits the ability to perform a frequency decomposition of a spike train. 
This model is a biologically plausible candidate for interpreting the spike trains typically observed in \gls{pc} afferents. 

We first demonstrate how a \acrfull{spll} can be designed using simple neurons and synapses, and subsequently discuss the properties that emerge when considering multiple spiking phase-locked loops with heterogeneous oscillations. 

We then illustrate how these phase-locked loops can be used to decode the spectral footprint of a given texture, while also showing that the model's response resembles recordings from a Macaque's cortex. 

Furthermore, we show how the frequency decomposition performed by the model is leveraged to outperform current \glspl{dsnn} in a classification task.

\section{Results}
\subsection{sPLL Analysis}
We propose here a \gls{snn} model, the \acrfull{spll}, that interacts with the spectral footprint of sequences of input spikes, instead of using the average interspike interval. The model comprises two building blocks, the \acrfull{tde} and the \acrfull{cco}, interacting in a closed loop. A detailed analysis of the model blocks is presented in Sec.~\ref{subsec:spll}. The model is depicted in Figure~\ref{fig:spll}(A). 

In this section we present the \gls{spll} response to an input spike train with a fixed rate (i.e., single input frequency), and we demonstrate that our model successfully locks to the input frequency. The spiking activity of the two \gls{spll} building blocks is illustrated in Figure~\ref{fig:spll}(B). The frequency sensitivity of the \gls{spll} is determined by the current $I_{\mathrm{bias}}$.

The three spiking neurons represent the INPUT, the \gls{tde} response and the \gls{cco} response. The frequency of the \gls{tde} is proportional to the phase difference between the spikes of the INPUT and the spikes of the \gls{cco}. The latter's frequency is determined by a stable input current $I_{\mathrm{bias}}$ as well as by the synapse connected to the \gls{tde}. The spikes of the INPUT and \gls{cco} exhibit a tonic spike regime (i.e., the spikes are evenly distributed across time), while the \gls{tde} spikes in bursts (i.e., the spikes are concentrated in a small period of time).

In Figure~\ref{fig:spll}(C), the traces of the synapses connecting the INPUT, the \gls{cco}, and the \gls{tde} are depicted. As mentioned in Sec.~\ref{subsec:spll}, the \gls{fac} and \gls{trg} do not integrate multiple stimuli ($f_{CCO} \ll 1/\tau_{\mathrm{TRG}}$, $f_{IN} \ll 1/\tau_{\mathrm{FAC}}$), while the \gls{syn} is integrating multiple spikes ($f_{TDE} \approx 1/\tau_{\mathrm{SYN}}$). This allows the two components of the \gls{spll} to work in two different regimes. The \gls{tde} responds to one combination of input spikes per time (one from the \gls{fac}, one from the \gls{trg}) using multiple spikes, coding for the time difference between them. The \gls{cco} integrates multiple spikes coming from the \gls{tde} but outputs a single spike.    

Figure~\ref{fig:spll}(D) shows the potential well visualization, which illustrates how the \gls{spll} locking mechanism works. 
In the example at hand the \gls{spll} responds to an input frequency of \SI{36}{\hertz} and an input current with magnitude $40$ (normalized current). 
The red and black lines indicate the case where the feedback is present or absent respectively (i.e., the \gls{tde} does or does not communicate with the \gls{cco}). 

In the figure we can see that, when the feedback is off, the phase is integrated periodically. 
This causes the \gls{tde} to respond with alternating high and low firing rate, which does not effectively encode information about the input firing rate.
This is illustrated by the black line passing between different level of shaded background, indicating different \gls{tde} spiking activity. 
By contrast, when the feedback is on, the \gls{tde} response is fed back to the \gls{cco}. 
As a result, a local minimum (i.e., a stable state) emerges in the potential landscape, which allows the system to effectively lock to the input frequency. 

Intuitively, when $\phi<0$, the inherent frequency difference forces $\phi$ to increase towards $0$, whereas if $\phi>0$, the spikes from the \gls{tde} cause the \gls{cco} frequency to increase, decreasing $\phi$ towards $0$. 
It is evident that regardless of the starting point, the system will eventually reach this state. 

Note that, however, since the \gls{tde} behaves complementary to the phase difference (its activity is maximal when $\phi=0$), the observed behavior is not that of a pure stable state at $\phi=0$ (as would happen in a typical \acrshort{pll}). The \gls{spll} alternates between being slightly above and slightly below $\phi=0$, as depicted by the system transitioning between high \gls{tde} spike counts and no \gls{tde} spikes in Figure~\ref{fig:spll}(B).

\begin{figure*}[tbh]
 \centering
 \includegraphics[width=\linewidth]{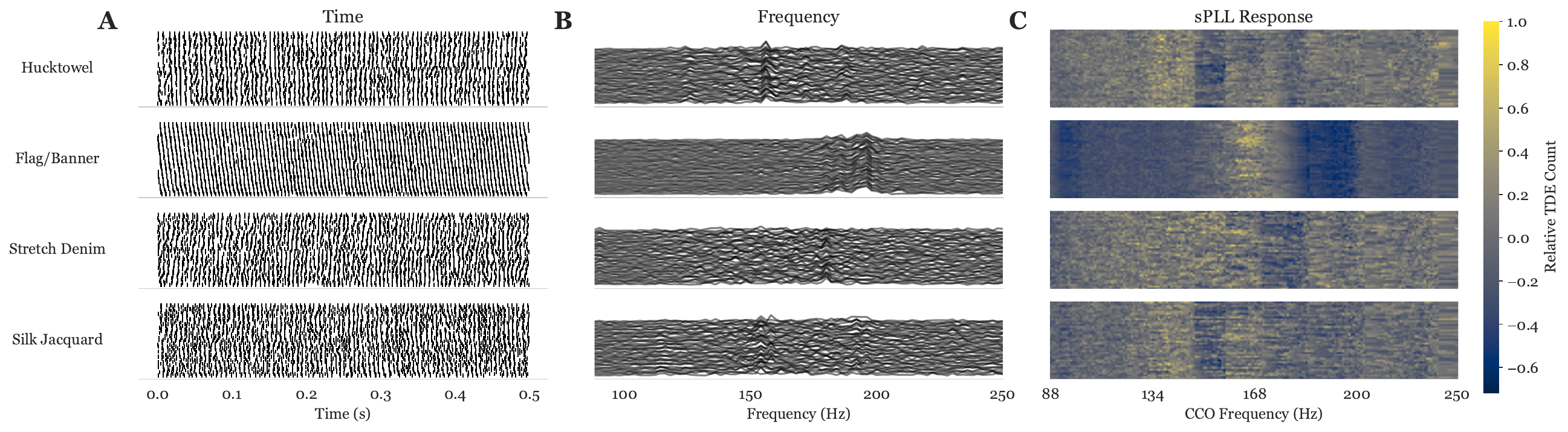}
 \includegraphics[width=\linewidth]{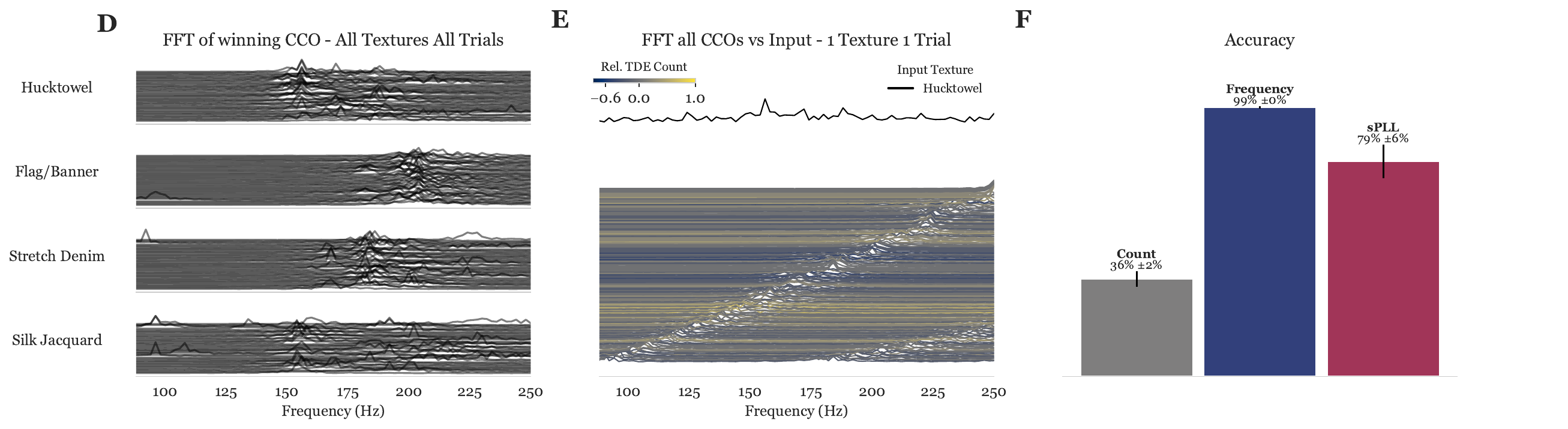}
  \caption{\gls{spll} layer processing data from \gls{pc} afferents. Panels \textbf{A} and \textbf{B} reproduce the plots depicted in Figure 2 of~\cite{Weber_Saal_ea2013}. The plots consists of real recordings of \gls{pc} afferents, obtained from a Macaque sliding its finger on different textures and kindly provided by the authors. Panel \textbf{A} is a raster plot while panel \textbf{B} is a normalized \gls{ft}. The different lines indicate different trials. Panel \textbf{C} illustrate the response of a layer of \gls{spll} to the recorded spikes, coded in overall spike counts of the \glspl{tde}. Panel \textbf{D} depicts the \gls{fft} curves for the \gls{cco} in the \gls{spll} layer with the lowest \gls{tde} spiking rate in panel \textbf{C}. The \gls{spll} with the lowest activity has a spiking frequency consistent across trials. In panel \textbf{E} a normalized \gls{fft} of a selected input spike train for the texture `Hucktowel' has been plotted. It is easy to observe that the spectral footprint of the `Hucktowel' texture (depicted as a black line at the top of the figure) has a peak around \SI{160}{\hertz}. The lower part of this panel shows the \gls{fft} of all the \glspl{cco} in response to the input spikes for `Hucktowel'. The color is derived from panel \textbf{C}, where each \gls{cco} is linked to its \gls{tde}. Panel \textbf{F} summarizes the test accuracy for a classifier trained on the spike count of the input spikes, the frequency footprint of the input spikes and the output activity of the \glspl{tde}. As we can see, while the spike count does not help in decoding the texture, the \gls{spll} spike count and the \gls{fft} exhibit similar performance.}
  \label{fig:bio}
\end{figure*}

\subsection{A Layer of sPLLs}
Given the \gls{spll}'s ability to lock onto specific frequencies, we can leverage this paradigm to detect the main frequency of a spike train. To illustrate this, we constructed a layer of 50 \glspl{spll} where each element has a distinct $I_{\mathrm{bias}}$ (ranging from \numrange{12}{60}, normalized current). When we feed an input spike train at a specific intrinsic frequency, the various \glspl{spll} respond differently. This is evident in Figure~\ref{fig:spll}(E1-E3), where the dots represent the instantaneous rate for every individual spike. The black dots represents the input spike train, while different colors represent different \glspl{spll}.

The three input stimuli (\SI{30}{\hertz}, \SI{33}{\hertz}, \SI{36}{\hertz}) interact differently with various \glspl{cco} based on the \gls{cco}'s intrinsic frequency (as defined by its input current $I_{\mathrm{bias}}$). \glspl{cco} that do not lock periodically change their frequency, following the phase accumulation between their intrinsic frequency and the input frequency (similar to what occurs in the potential well example in Figure~\ref{fig:spll}(D) when no feedback is present). This behavior is visible in Figure~\ref{fig:spll}(E1), where higher frequencies display a periodic behavior dictated by phase accumulation.

The \glspl{spll} that manage to lock to the input do not accumulate phase. This results in a stable phase difference and consequently a more regular \gls{tde} spike activity. It's noteworthy that several \glspl{cco} can lock to an input frequency due to the feedback loop that adjusts the internal frequency. This can be understood by examining the spread of the potential well in Figure~\ref{fig:spll}(D) for the case with active feedback.

The distinct activity of the \gls{tde} can be utilized to easily identify the \glspl{spll} that managed to lock more effectively. This is apparent in Figure~\ref{fig:spll}(F), where the overall spike count of the \gls{tde} is plotted versus input frequency and versus \gls{cco} intrinsic frequency. The spike count of the \gls{tde} when receiving an input spike train was normalized over the input stimulus and over the intrinsic frequency of the \gls{cco} for clarity. The normalization can be justified by considering that a \gls{wta} layer could be placed at the output of the \gls{spll} network. The plot demonstrates that for any given input frequency, compatible with the investigated range of \gls{cco} intrinsic frequencies, there is an \gls{spll} that is able to lock, leading to a low \gls{tde} spike count. The intrinsic frequencies of the \gls{cco} (the ticks of the x-axis in Figure~\ref{fig:spll}(F)) were estimated by allowing the \gls{spll} to run without an input stimulus (so the \gls{cco} was not excited by the synapse but only by the $I_{\mathrm{bias}}$).

Another observation in Figure~\ref{fig:spll}(F) is that, close to each locking \gls{spll}, there is another \gls{spll} that spikes intensively. This is because when two spike trains have a low frequency difference, but not low enough to lock, or only lock poorly, the \gls{tde} spends a lot of time residing in the region where the firing rate is high, as visible in Figure~\ref{fig:spll}(D).

Additionally, Figure~\ref{fig:spll}(F) highlights the harmonics of the input signals. The \gls{spll} doesn't only lock with the input frequency but also with its multiples. This is mainly due to different harmonics of spike rate not being discernible because lower frequencies could be interpreted as higher frequency spike trains with missing spikes.

\subsection{Modeling the somatosensory pathway}
As explained in the introduction, the hypothesis of the existence of a phase-locked loop in the somatosensory pathway was proposed by Ahissar in the late '90s~\cite{Ahissar1998}. 
The model presented there identified the cortex as the location of an oscillator, coupled with a phase detector in the thalamus. 

 \begin{figure*}[tbh]
 \centering
 \includegraphics[width=\linewidth]{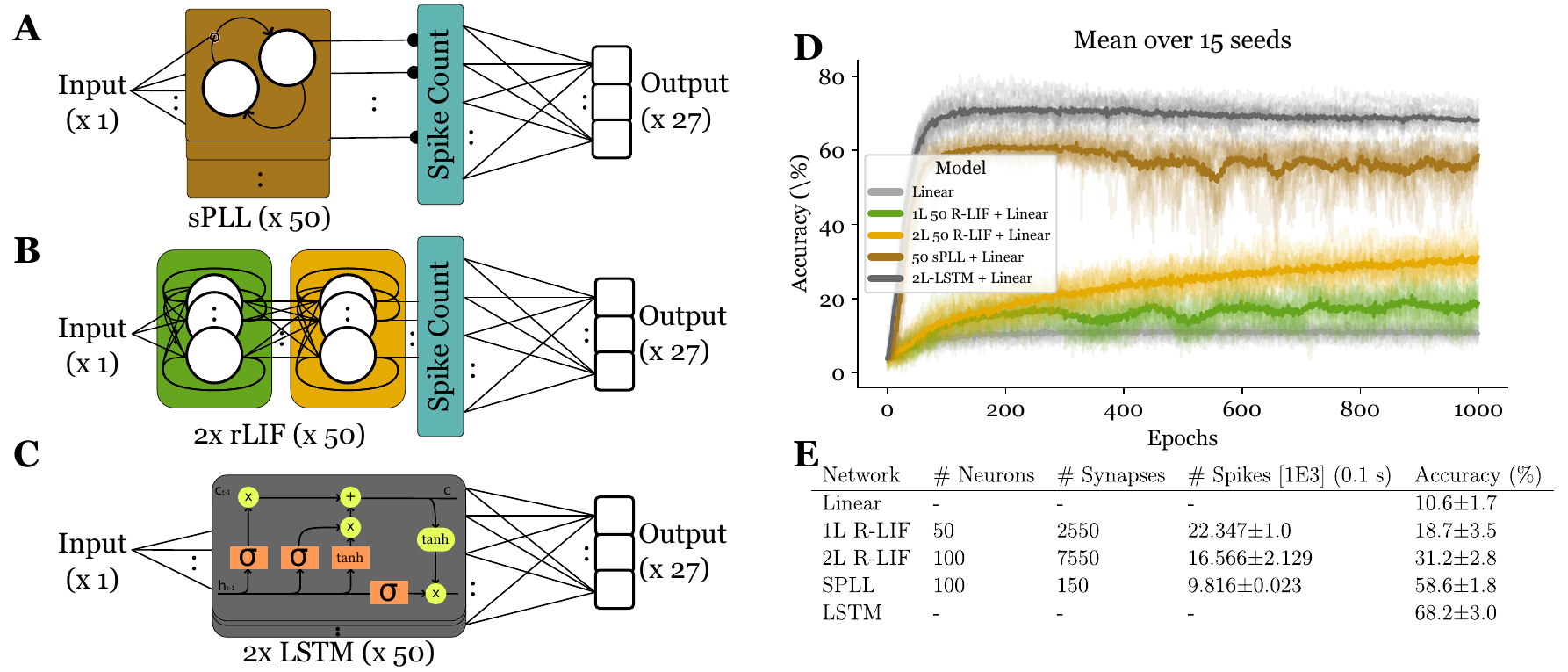}
 \caption{In scheme \textbf{A}, \textbf{B}, \textbf{C} the four networks under test: 1 layer of 50 \gls{spll}, 1 and 2 layers of 50 \gls{r-lif}, and a \gls{lstm} composed of 2 layers with output dimensionality of 50. All the networks are then passed through a linear classifier. In panel \textbf{D} the test accuracy over training of the four networks. The networks were tested against a linear classifier applied to the input spike count. The plot is composed of 15 different instances per network, with the mean plotted over it. In table \textbf{E} is shown the comparison among the different networks in terms of accuracy and in terms of resources (number of synapses and total spikes). The \gls{lstm} figure has been adapted from~\cite{Chevalier23} under license CC BY-SA 4.0.}
 \label{fig:multifreqlabel}
\end{figure*}
The \gls{spll} described in Sec.~\ref{subsec:spll} provides novel hypotheses about the biological somatosensory pathway. 
To quantitatively assess the similarity between the response of our model and biological data, we compare the \gls{spll} response to in-vivo recordings of peripheral~\cite{Weber_Saal_ea2013} and cortex~\cite{Lieber_Bensmaia2019a} data from rhesus macaques.

In Figure~\ref{fig:bio} (A and B) we recreated the peripheral recording depicted in Figure~2 from~\cite{Weber_Saal_ea2013}, using the \gls{pc} response data kindly provided by the authors. 

The figure shows data for 4 textures, with 63 trials each. 
The first part of the Figure illustrates the raster plots of the \glspl{pc} response (Figure~\ref{fig:bio} (A)). 
Despite the nontrivial task to discriminate textures based on the raster plots (e.g. stretch Denim vs silk Jacquard) by visual inspection, it has been demonstrated that macaques are able to discriminate different textures of this kind when sliding their finger on them~\cite{Long_Greenspon_ea2022}. 

To characterize the frequency components of the \gls{pc} response, we performed an \gls{ft} analysis  to extract the power spectrum from the neural recording data (see Fig.~\ref{fig:bio} (B)). 
One can qualitatively observe that, in contrast to the raster plots, the power spectrum of the spikes clearly provides footprints for different textures, with very low trial-to-trial variability. 

To numerically assess this qualitative observation, the authors in~\cite{Weber_Saal_ea2013} studied the correlation of both the mean firing rate and power spectra data with the input textures, demonstrating the superiority of frequency-based methods in decoding fine textures.

The similarity between our model and the afferent pathway projecting from the peripheral nerves was then studied by looking at the response of the \gls{spll} to \gls{pc} spiking activity. 
The experiment employed 200 \glspl{spll} in parallel, all fed with different currents, driving the different \glspl{cco} in a range between \SI{50}{\hertz} and \SI{125}{\hertz}. 
The biological data from~\cite{Weber_Saal_ea2013} were presented as input for 2 seconds. The response of the \gls{tde} was then recorded over the experiment for all the different \glspl{spll}. 

In Figure~\ref{fig:bio}(C), we report the results of the experiment. 
The x-axis depicts the frequencies at which the different \glspl{cco} are spiking, while the y-axis reports different trials. 
The color indicates the total spike count of the \glspl{tde} during the experiment.
A normalization has been applied to acount for the intrinsic frequency difference between \glspl{spll}.
This normalization model fit a line to the intrinsic dependence between the $I_\mathrm{bias}$ and the \gls{tde} spike count, removing it for a fair comparison between \glspl{spll}.

The \gls{spll} response demonstrates a clear texture selectivity and low trial-to-trial variability. 
The locking behavior of the \gls{spll}, as explained in Section~\ref{subsec:spll}, is expressed by regions of reduced response (shown in orange in Figure~\ref{fig:bio}(C)) surrounded by enhanced response (in purple). 
These regions are closely related to the power spectrum peaks present in the \gls{pc} activity (Figure~\ref{fig:bio}(B)). 

In order to quantify the performance of the \gls{spll} model at detecting frequencies, we trained a linear classifier to predict the input texture on the basis of the data illustrated in Figure~\ref{fig:bio}(A-C). 

The classifier was trained on three different datasets. 
The first was the spike count in Figure~\ref{fig:bio}(A), where the network first randomly projected to 200 neurons and then a linear classifier decoded it. 
The second was instead tested on the \gls{fft} of the spikes with a window of \SI{200}{\hertz}. The third was trained using the outputs of the 200 \glspl{spll}. 

Figure~\ref{fig:bio}(F) shows that the accuracy of the classifier is low when using the \gls{pc} spike count (\SI{36}{\percent}), while it is high when using the power spectrum (\SI{99}{\percent}). The \gls{spll} response produces an accuracy closer to that of the \gls{fft}, demonstrating that it can extract the spectral footprint (\SI{81}{\percent}).
This result suggests that the \gls{spll} model is a suitable candidate for the identification of the neural substrate of texture perception.

In Figure~\ref{fig:bio}(D-E) we show the behaviour of the \glspl{cco} in response to different inputs, in order to compare it to what biological literature reports~\cite{Lieber_Bensmaia2019a}. 

Starting from Figure~\ref{fig:bio}(E), the \gls{fft} of a single trial of the Hucktower texture is presented, along with the \glspl{fft} of 200 different \glspl{cco}. 
In the top part the Hucktower texture \gls{fft} is shown as a single black line, while the rest of the plot is occupied by the \gls{fft} of the \glspl{cco} responding to the texture spikes. Every \gls{fft} is normalized with respect to its maximum and shifted according to the index of the \gls{cco} to have an informative view of the different trials. The color codes the overall spike count for the corresponding \gls{tde}, extracted from Figure~\ref{fig:bio}(C), where blue indicates low activity, and yellow indicates high activity. Low activity in the \gls{tde}, usually associated with the locking behaviour, can be found in the frequency range comparable to the input frequency peak (around \SI{80}{\hertz}) and in the harmonics associated with the peak (around \SI{160}{\hertz}). This plot suggests that the \gls{tde} activity reflects the position of the \gls{cco} with the behaviour closest to the input frequency. This can be used to identify the frequencies at which the \gls{pc} afferents spike, thanks to the locking of the \gls{cco}. 

Figure~\ref{fig:bio}(D) shows the \gls{fft} of the \gls{cco} associated with the \gls{tde} with the lowest spiking activity for multiple inputs and trials, following the method as used for Figure~\ref{fig:bio}(E). One can see that, for a given texture, the behaviour of the \glspl{cco} is associated with the spectral footprint of the input. This result is in accordance with the experimental observation published in~\cite{Lieber_Bensmaia2019a}. In said paper, cortex neurons projecting from \gls{pc} afferents exhibit very consistent spectral footprints in response to input stimuli across trials, as visible in the reference's Figure~3(E) in the supplementary material.   

The results presented here suggest that the developed model of the \gls{spll} has several points in common with the biological recordings performed in~\cite{Weber_Saal_ea2013}
and~\cite{Lieber_Bensmaia2019a}. First we demonstrated that our model is able to decode the frequency footprint present in the spikes more efficiently than a simple spike count and with results comparable to an \gls{fft}. After this, we showed, in Figure~\ref{fig:bio}(D), that the behaviours exhibited by components of our models can be similar to those demonstrated in the biological counterparts. 

\subsection{Multiple frequency decoding}
After having shown that the \gls{spll} can decode complex spectral footprints from biological spike trains, we explore the possibility of using this ability to perform computation in artificial agents, where complex spike patterns could be used to carry more variables in the same spike train. This possibility is discussed in Section~\ref{subsec:isivsft} in the Supplementary Material, where \gls{mst} are decoded through the \glspl{isi} and \glspl{fft}.

Regarding the implementation of architectures for decoding \gls{mst}, ~\cite{Jimenez-Fernandez_Cerezuela-Escudero_ea2017}  proposed an algorithm that processes spike trains using a \gls{slpf} bank emulating the cochlea's frequency decomposition. The \gls{slpf} is composed of a closed loop architecture made of a \gls{shf}, a \gls{sig} and a \gls{sfd}. The output of each \gls{slpf} is a spike train where the spikes of a specific frequency have been removed. This algorithm therefore returns a bank of spike trains, in which each bank represents the spiking activity of the input, decomposed in a specific frequency range. The model used here, despite using spikes, heavily bases its functionality on clocked signals, due to the way it handles the individual events. This is a drawback when handling sparse time series as spike trains because it makes difficult to harness the low power computation of event-driven design. 

To test the ability of the \gls{spll} to detect multiple frequencies in a spike train, we performed a classification task where the inputs were composed of complex spike trains with two nested frequencies in it. The first frequency ($F_1$) was swept between \SI{30}{\hertz} and \SI{54}{\hertz}, while the second frequency was swept between \SI{30}{\hertz} and \SI{57}{\hertz} as depicted in Figure~\ref{fig:accuracy_comparison_isi_fourier}(A) (explained in the Methods). Using this dataset, we compared our architecture with other notable architectures. 

We first built a network composed of 50 \glspl{spll} with progressive $I_{\mathrm{bias}}$ and shared input. The input was connected to the \gls{fac} terminal of each \gls{spll}, as visible in Figure~\ref{fig:multifreqlabel}(A). The output of the \gls{spll} network was then recorded over time and the total spikes during the simulation of each \gls{tde} were counted. This spike count was then used to train and test a linear classifier. 

We then created two \gls{dsnn} networks of r-\glspl{lif} as explained in Sec.~\ref{subsec:lif_network}. One network was only composed of 50 neurons, connected in a one-to-all fashion to the input. They were then recurrently connected to the other neurons of their layer. Their spikes were counted and the resulting vector used for training a linear classifier. The training was acting both on the weights of the linear classifier as well as on the weights of the r-\glspl{lif} network. In the case of two layers of neurons, a second layer of 50 r-\glspl{lif} was introduced, as visible in Figure~\ref{fig:multifreqlabel}(B). In this case the training modified also the weights of the second layer.

To define the low and high boundaries of the accuracy we introduced two other networks: a simple linear classifier and a \gls{lstm}. The \gls{lstm} was composed of 2 layers with an output dimensionality of 50. The hidden state of the \gls{lstm} $h$ was used as input to a linear classifier. We found that the simple linear classifier applied to the spike count of the input was not enough to discern different frequencies whereas the \gls{lstm} (Figure~\ref{fig:multifreqlabel}(C)) showed a very good accuracy.

The results of the simulation are presented in Figure~\ref{fig:multifreqlabel} (plot D and table E). The former depicts the test accuracy over epochs, while the latter summarizes the figures of merits. As we can see, the \glspl{spll} layer performs closely to the \gls{lstm}, while the two networks made of \gls{r-lif} struggle to detect the different frequencies. 

This is especially striking when comparing the resources. The \gls{spll} is composed of two neurons and three synapses per element. This means that for the overall network there were 100 neurons and 150 synapses. The single layer \gls{r-lif} is composed of 50 neurons, where each neuron is connected to the input and to the other neurons in the layer for a total of 2550 plastic synapses. For the same reason, The 2 layer \gls{r-lif} has 100 neurons and 7550 plastic synapses. 

Furthermore, the number of spikes generated by the \gls{spll} is lower than the two \glspl{dsnn}, as can be seen in Figure~\ref{fig:multifreqlabel}(table E). 
This is due to the fact that the \gls{spll} needs few spikes to abstract the overall spectral footprint of the stimulus, while the \gls{dsnn} requires more spikes to capture the temporal dynamics, and thus to be able to effectively decode the proposed input. 
The number of spikes has been chosen as a metric here because in artificial agents it often dictates the power consumption. This is due to the fact that the event generated by the spike is responsible for triggering a change, increasing the power consumption over time. Typical energy measures for such events are in the order of \unit{\femto\joule}-\unit {\nano\joule} in neuromorphic processors~\cite{Davies_Srinivasa_ea2018a,Frenkel_Legat_ea2019,Mastella_Toso_ea2020}, giving us a reasonable estimate of the power consumption of one inference. Besides the lower overall spike count, the standard deviation for the total spike count is lower in case of the \gls{spll} as well, which can be attributed to the fact that for the \gls{dsnn} the full network was trained with backpropagation, whereas in the \gls{spll} only the linear classifier at the end was part of the training.

This result is striking because it shows that our model can accurately detect the frequencies in a multi-frequency signal better than complex \glspl{snn}  composed of recurrent connections and trained using backpropagation, while maintaining a low and fixed connectivity pattern and sparse firing activity.

Importantly, it highlights how recurrent \gls{dsnn} are more suited for simple spiking patterns and are not fit for the case where a spike train is composed of multiple elements. 

\section{Conclusion}
In this paper, we presented a novel model for a spiking neural network, the \acrfull{spll}, inspired by a previous work on phase-locked loops in the somatosensory cortex. We illustrated how the model works in simulation, highlighting how the \gls{spll} manages to lock to an input frequency and how to use this behaviour to decode input frequencies. 

We then presented the similarity between our model and biological data, hypothesizing that we proposed a good candidate for modeling the brain's micro-circuitry: the \gls{spll} is able to decode the texture represented by spike trains recorded from peripheral nerves in-vivo (Macaque data \cite{Lieber_Bensmaia2019a}), and the activity of the oscillator inside the model (the \gls{cco}) resembles recordings from cortical neurons projecting from \gls{pc} afferents. 

Moreover, we illustrated our model's capability to decode multiple frequencies contained within a complex spike train. This feature positions the model as a promising candidate for advancing artificial touch and endowing artificial agents with unprecedented capabilities.

For future works, the possible paths to take are several, both towards improved \glspl{snn}, biological resemblance, and real-life applications.

We showed that our model functions as an essential component within a \gls{snn}, decoding frequency components in spike trains. In this regard, our model can be seamlessly integrated as a layer within the \gls{snn} architecture. A PyTorch module of the model is available for this purpose in the public repository at \href{https://github.com/bics-rug/sPLL}{\url{github.com/bics-rug/sPLL}}

Also, the \gls{spll} paradigm could be harnessed for a novel method of information transmission across layers: a single spike train can convey multiple variables on multiple channels, distinguished by differing frequency bands, reminiscent of frequency-modulation encoding. Sets of \glspl{spll} could exhibit sensitivity to specific frequency ranges and be able to capture information within these ranges, thereby facilitating a more streamlined communication channel.

Regarding biological resemblance, our model could have a deeper level of similarity to biology than what was explored here. A key feature of texture perception in biology is the independence from sliding speed. Our model can naturally be extended to exhibit this property. In fact, the $I_{\mathrm{bias}}$ current that defines the intrinsic frequency of every \gls{cco} can be adjusted, moving the selectivity of frequencies of a given \gls{spll} to a different range. In a closed loop system with proprioception, this current can be dynamically modulated by the sliding speed. This approach  would be coherent with the scheme proposed in~\cite{Ahissar_Oram2015}, where the motor cortex (responsible for the sliding velocity) is projecting to the somatosensory cortex, informing it about the current strength applied to the muscles. 

Another important extension of this work is to make our model compatible with applications relevant for artificial agents. In particular, an artificial agent used to assess the quality of fabrics may be given the task of decoding the perceived texture. In this scenario, data would be converted from analog pressure signals into spikes using transducers equivalent to the \gls{pc} corpuscles, and the motor information would then be fed to the \glspl{spll}. 

For this purpose, a \gls{cmos} circuit could be developed to decode tactile stimuli. Implementing this would be straightforward, given the documented existence of \gls{cmos} equivalents for all the constituent building blocks~\cite{Bartolozzi_Indiveri07,Livi_Indiveri09,Chicca_Stefanini_ea2014,Milde_Bertrand_ea2018}. In this scenario, the integrated circuits could be embodied into robots or prosthetics, enhancing the ability of artificial agents to discern textures, alongside other sensory stimuli such as vision and audio~\cite{Ahissar_etal23}, while maintaining extremely low power consumption and minimal latency~\cite{Bartolozzi_etal16}. Currently, a realized \gls{cmos} circuit implementing these functions has been built and tested but is yet to be made publicly available.
\section*{Acknowledgment}
We would like to thank the following individuals for their contributions to this research: Nicoletta Risi and Hugh Greatorex for their thorough revision and improvement of the work. Giulia Corniani, Miguel Casal Santiago, Hannes Saal, and Steven Abreu for their insightful discussions that enhanced the project. Justin Lieber for providing the datasets from~\cite{Weber_Saal_ea2013,Lieber_Bensmaia2019a}, along with insights about them. Sliman Bensmaia and his group, whose work has inspired this research. Matthew Cook for his help with the manuscript. These contributions have been invaluable in shaping the content and direction of this work. The authors would like to acknowledge the financial support of the CogniGron research center and the Ubbo Emmius Funds (Univ. of Groningen). This work has been supported by EU H2020 projects NeuTouch (813713). We thank the Center for Information Technology of the University of Groningen for their support and for providing access to the Hábrók high performance computing cluster.

\section{Models}
\label{sec:models}
In this section we introduce the computational models used in this work. We first talk about the novel network proposed here, the \acrfull{spll}, along with the parts of which it is composed: the \acrfull{cco} and the \acrfull{tde}. After that, we talk about the \acrfull{dsnn} and the \acrfull{lstm}, used to compare the novel network with \gls{sota}.

\subsection{Current Controlled Oscillator}
The \acrfull{cco} is an oscillator that changes its frequency in relation to an input current. In this work, this is realized with a \acrlong{cuba} \acrlong{lif} (\acrshort{cuba}-\acrshort{lif}), comprised of an integrating synapse and an integrating neuron.

The synapse can be represented by the following differential equation:
\begin{equation}
 \frac{dI_{\mathrm{syn}}}{dt} = \frac{-I_{\mathrm{syn}}+\sum_j\mathrm{gain}_{\mathrm{syn}}\cdot \mathrm{spk}_{\mathrm{in}}^j}{\tau_{\mathrm{syn}}}
\label{eq:synapse}
\end{equation}
where:
\begin{itemize}
 \item $I_{\mathrm{syn}}$ is the internal state of the synapse;
 \item $\mathrm{gain}_{\mathrm{syn}}$ is the voltage-to-current conversion factor;
 \item $\mathrm{spk}_{\mathrm{in}}$ represents the input spike (voltage);
 \item $\tau_{\mathrm{syn}}$ is the synapse time constant
\end{itemize}

Upon receiving a spike from a neuron, the synapse's internal variable (representing a current) increases by $\mathrm{gain}_{\mathrm{syn}}/{C_{\mathrm{syn}}}$, and slowly decreases with a time constant of $\tau_\mathrm{syn}$. 

The relationship between the input spike rate and the reached maximum current in this first-order differential equation can be obtained by considering the case in which a spike arrives either before or after the internal variable of the synapse is completely discharged. The formula for this would be:
\begin{multline}
I_{\mathrm{syn}}^\mathrm{max}\approx
\begin{cases} 
\frac{\mathrm{gain}_{\mathrm{syn}}}{C_{\mathrm{syn}}} & \mathrm{if}\ f_{in} \ll \frac{1}{\tau_{\mathrm{syn}}}\\
\frac{\mathrm{gain}_{\mathrm{syn}}}{C_{\mathrm{syn}}}Ne^{-\frac{1}{f_{in}\tau_{\mathrm{syn}}}} & \mathrm{else} 
\label{eq:synapse_max}
\end{cases} 
\end{multline}
where $f_{\mathrm{in}}$ is the frequency of the input and $N$ is the number of spikes that have occurred ($T_{\mathrm{sim}}$/$f_{\mathrm{in}}$). This means that if the internal variable of the synapse was not fully discharged before the arrival of the next spike the internal variable sums the contribution of multiple spikes. 

The neuron, specifically a \gls{lif} model, follows a similar equation to that of the synapse, with the difference being that when the variable $V$ exceeds a threshold, it resets to $0$ for a refractory time $T_{\mathrm{refr}}$:
\begin{multline}
\begin{cases} 
\frac{dV_{\mathrm{mem}}}{dt} = -\frac{V_{\mathrm{mem}}}{\tau_{\mathrm{neu}}} + \frac{I_\mathrm{in}}{C_{\mathrm{mem}}} \\
\mathrm{if}\ V > V_{\mathrm{thr}}\ \&\ \mathrm{count_{refr}} = 0 \rightarrow \mathrm{spk}\\
\mathrm{if}\ \mathrm{spk} \rightarrow \mathrm{count_{refr}} = T_{\mathrm{refr}},\ V=0 \\
\mathrm{if}\ !\mathrm{spk} \rightarrow \mathrm{count_{refr}} = \mathrm{count_{refr}} - 1
\end{cases} 
\end{multline}
where $I_{\mathrm{in}}$ is the input current fed to the neuron, which can originate from a synapse or a stable current input:
\begin{equation}
 I_{\mathrm{in}} = I_{\mathrm{syn}} + I_{\mathrm{bias}}
\label{eq:current_decomposition}
\end{equation}
Assuming a constant current at the input, and solving the first-order differential equation, we can estimate the time when the neuron reaches the threshold as:
\begin{equation}
\begin{cases} 
 t_{\mathrm{thr}} = \tau_{\mathrm{neu}}\ln\left(\frac{I_{\mathrm{in}}}{I_{\mathrm{in}} -\frac{C_{\mathrm{mem}}}{\tau_{\mathrm{neu}}}V_{thr}}\right) \\
  t_{\mathrm{thr}} =  T_{\mathrm{refr}} & \mathrm{if\ t_{\mathrm{thr}} >  T_{\mathrm{refr}}} 
 \end{cases}
\end{equation}
Given that the frequency of a neuron is defined as the average \gls{isi}, we can say that $f_{\mathrm{neu}} = 1/t_{\mathrm{thr}}$. Note that the frequency of the neuron is capped by the refractory period. By imposing that the neuron is controlled by a constant current, we have thus created a \gls{cco} using a simple neuron.

\subsection{Time Difference Encoder}
The \acrfull{tde} is a model used in neuromorphic computing to translate the time difference between two input spike trains into a spike rate~\cite{Milde_Bertrand_ea2018,DAngelo_Janotte_ea2020,Schoepe_Janotte_ea2021,Gutierrez-Galan_Schoepe_ea2021}, which consists of a gated synapse and a neuron. 
The gated synapse is composed of two traces, \gls{fac} and \gls{trg}, as illustrated in Figure~\ref{fig:spll}, where the gating is graphically represented with a ring around the \gls{trg}. Both \gls{fac} and \gls{trg} are similar to the trace presented in Eq.~\ref{eq:synapse}, except for the modulation of \gls{trg}'s $\mathrm{gain}_{\mathrm{syn}}$ by \gls{fac}. 
\begin{multline}
\begin{cases}
 \frac{dI_{\mathrm{FAC}}}{dt} = -\frac{I_{\mathrm{FAC}}}{\tau_{\mathrm{FAC}}} + \frac{{\mathrm{gain}_{\mathrm{FAC}} \mathrm{spk}_{\mathrm{FAC}}}}{C_{\mathrm{FAC}}} \\[10pt]
 \frac{dI_{\mathrm{TRG}}}{dt} = -\frac{I_{\mathrm{TRG}}}{\tau_{\mathrm{TRG}}} + \frac{{\mathrm{gain}_{\mathrm{TRG}}I_{\mathrm{\mathrm{FAC}}}\mathrm{spk}_{\mathrm{TRG}}}}{C_{\mathrm{TRG}}}
\end{cases}
\end{multline}
As demonstrated in~\cite{Greatorex_etal24}, we can estimate that the current of the \gls{trg} depends on the time difference between the spikes' arrival time from the two channels (defined here as $t_{TRG}$ and $t_{FAC}$) and on the gain of both the \gls{fac} and the \gls{trg}.
\begin{align}
I_{\mathrm{{TRG}}}(t) = 
\begin{cases}
 I_{\mathrm{gain}}e^{-\frac{\Delta t}{\tau_{\mathrm{FAC}}}}e^{-\frac{t-t_{\mathrm{TRG}}}{\tau_{\mathrm{TRG}}}} & \mbox{if $\Delta t > 0$} \\[6pt]
 0 & \mbox{else}
 \end{cases}
\label{eq:tde_final}
\end{align}
where $\Delta t = t_{\mathrm{TRG}}-t_{\mathrm{FAC}}$ and $I_{\mathrm{gain}} = I_{\mathrm{gainFAC}}I_{\mathrm{gainTRG}}$.

The \gls{trg} current is then fed into a \gls{lif} neuron, which responds with spikes related to the time difference between the \gls{fac} and the \gls{trg} channels. 
In the end, assuming a logarithmic dependence between the spiking activity and the current, and an exponential dependence between the current and the time difference, the frequency of the \gls{tde} is given by:
\begin{align}
 f_{\mathrm{tde}} \propto \frac{1}{t_{\mathrm{TRG}}-t_{\mathrm{FAC}}}
 \label{eq:tde_relation}
\end{align}
Note that this last approximation assumes that the \gls{tde}'s traces are at their minimum every time an input spike arrives at both the \gls{trg} and \gls{fac}. This is possible if the input frequencies are lower than the reciprocal of the model's $\tau$s ($f_{in} \ll 1/\tau_{FAC},1/\tau_{TRG}$), as visible in Eq.~\ref{eq:synapse_max}. 

\subsection{Spiking Phase-Locked Loop}
\label{subsec:spll}
This model, known as the \acrfull{spll}, was initially introduced in~\cite{Mastella_Chicca2021}, inspired by previous works in~\cite{Ahissar1998}. It comprises a \gls{cco} and a \gls{tde} arranged in a feedback loop, as depicted in Figure~\ref{fig:spll}(A).

The model functions as follows: the \gls{cco} receives a stable current from an external terminal ($I_{\mathrm{bias}}$), inducing spiking at a specific frequency. The \gls{tde} receives spikes from the input via the \gls{fac} terminal and from the \gls{cco} through the \gls{trg} terminal. Subsequently, the output of the \gls{tde} is channeled to a synapse that stimulates the \gls{cco}. This stimulation results in the adjustment of the operating frequency of the \gls{cco}. The variation in frequency is subsequently employed to minimize the phase differential between the input and the phase of the \gls{cco}.

As illustrated in Figure~\ref{fig:spll}(A), the spikes are integrated into currents using synapses (due to the \gls{cuba} property), which then feed into the subsequent neuron. This method is used convert voltage signals (the spikes) into current signals (the traces) and to give temporal kernels to the spikes.

The equation governing the \gls{spll} is an interacting version of the two models: the \gls{cco} and the \gls{tde}. As explained in the \gls{tde} section, if we consider that the frequency of the INPUT and the frequency of the \gls{cco} are both lower than the reciprocal of the \gls{tde}'s $\tau$s, we can utilize Eq.~\ref{eq:tde_relation}. Conversely, if we assume that the input frequency of the \gls{cco} is higher than the reciprocal of its $\tau$ but still lower than the reciprocal of the threshold period $\frac{1}{T_{\mathrm{refr}}}$, we can state that:

\begin{equation}
f_{\mathrm{cco}} \propto \ln\left(\frac{I_{in}}{I_{in} + \frac{C_{\mathrm{mem}}}{\tau_{\mathrm{neu}}}V_{\text{thr}}}\right) \approx \ln\left(\frac{I_{in}}{\frac{C_{\mathrm{mem}}}{\tau_{\mathrm{neu}}}V_{\text{thr}}}\right)
\end{equation}
Considering now that the \gls{tde} has a spiking rate comparable to $\tau_{\mathrm{SYN}}$ we can say that:
\begin{equation}
    I_{\mathrm{SYN}} = \frac{\mathrm{gain}_{\mathrm{syn}}}{C_{\mathrm{syn}}}Ne^{-\frac{1}{f_{in}\tau_{\mathrm{syn}}}}
\end{equation}
where N is the number of spikes emitted by the \gls{tde}.
So we have that:
\begin{equation}
f_{\mathrm{cco}} \propto \frac{ln\left(N\right)}{\tau}  \approx \frac{N}{\tau}  \approx f_{\mathrm{tde}}
\end{equation}

Now, assuming that the spike trains are periodic ($t_{\text{spike}} = \phi_{\text{spike}}$), we can further deduce:
\begin{align}
\begin{cases}
f_{\text{tde}} \propto \phi_{\text{cco}}-\phi_{\text{in}} \\
f_{\text{cco}} \propto f_{\text{tde}} \\
\phi_{\text{cco}} = \int f_{\text{tde}}
\end{cases}
\end{align}

Here, $\phi_{\text{cco}}$ and $\phi_{\text{in}}$ represent the phases of a given spike in the \gls{cco} and INPUT, respectively.
\subsection{Deep Spiking Neural Network}
\label{subsec:lif_network}
To compare the performance of the \gls{spll}, we opted for the most common neuromorphic algorithm for classification: a \acrfull{dsnn}. This network consists of several layers (either 1 or 2 hidden layers in this work) of \gls{lif} neurons interconnected using integrating synapses (\acrshort{cuba}-\acrshort{lif}).

In this architecture, each neuron communicates with the subsequent layer and within its own layer through an all-to-all connection, thus forming a \gls{rnn} of \gls{r-lif}. We utilized \gls{bptt} and surrogate gradients to train the input synapses of every layer, as well as the recurrent ones.

\subsection{Long Short-Term Memory}
We also employed an algorithm suited for temporal datasets to benchmark the \gls{spll} against the traditional \gls{sota}: \acrfull{lstm}. This model, first introduced by~\cite{Hochreiter_Schmidhuber1997}, represents an improvement over traditional \glspl{rnn}, effectively addressing the vanishing gradient problem.

The \gls{lstm} model, depicted in Figure~\ref{fig:multifreqlabel}(C), consists of a complex unit named a \textit{memory cell}, that stores the time sequence used for computation. It comprises three gates: the input gate, the output gate, and the forget gate. These gates play a crucial role in regulating the information stored in the memory cell. The forget gate decides which existing information in the memory cell should be discarded, the input gate determines which parts from the input should be retained, and the output gate defines which portions of the stored values should be propagated to the output.

In this work, we utilized a two-layer \gls{lstm} with an input dimensionality of 1 and an output dimensionality of 50. The internal state $h$ was used as a vector for training a classifier. The parameters of the \gls{lstm} were trained using backpropagation.

\section{Methods}
\label{sec:methods}
\subsection{Multifrequency Spike Trains}
\label{subsec:spikes}
Neurons in neuromorphic systems communicate using digital voltage pulses with fixed height and width, known as spikes. The information transmitted between neurons can be encoded in the time difference between these spikes. 

In the simplest case in which the time between spikes is constant over time, the spike train can be described by a Dirac comb:
\begin{equation}
 R(t) = \frac{1}{T} \sum_k {\delta(t-kT)}
 \label{eq:dirac_comb}
\end{equation}

To account for noise and variability, we consider the formula:
\begin{equation}
 R(t) = \frac{1}{T} \sum_k {\delta\left(t-\left(kT + \mu+\sigma[k]\right)\right)}
 \label{eq:noise}
\end{equation}
where $\mu$ is a random time shift that affects all the $\delta$s equally and
$\sigma$ represents jitter noise which modifies the time of each spike independently.

More generally, spike trains differ from Dirac combs in that they do not have to be periodic, and each spike can have an arbitrary distance from the previous one. Thus, we define:
\begin{equation}
R(t) = \sum_k {\frac{1}{\mathrm{ISI}_k}\delta(t-t_{s}[k])}
\label{eq:spike_train} 
\end{equation}
where $t_s[k]$ is the time of the $k-th$ spike. The \glspl{isi} are defined as the time differences between consecutive spikes: $\mathrm{ISI}_k=t_s[k]-t_s[k-1]$. In case the spike train is generated from a Dirac comb with the addition of jitter noise the time of the $k-th$ spike is given by $t_{s}[k]=kT +\mu +\sigma[k]$.

Extending from the definition of a spike train, we consider spike trains with multiple frequencies, referred to as \glspl{mst}. They are created by merging spike trains with different frequencies:
\begin{multline}
R(t) = \mathrm{merge}(R_0(t),R_1(t)) = \\ \frac{1}{T_1} \sum_{k} {\delta(t-t^{1}_{s}[k])}\ |\ \frac{1}{T_2} \sum_{k} {\delta(t-t^{2}_{s}[k])} 
\label{eq:merge_operation}
\end{multline}
where each train $R_i(t)$ has its period $T_i$ and the temporal jitter is sampled from a Gaussian random distribution, e.g. $t^{i}_{s}[k]=kT_i +\mu_i+\sigma_i[k]$.
The operation performed here, akin to an OR gate ($|$), merges the two spike trains into a single time series. Note that this means that when two spikes coincide, we consider them as a single spike, with the usual fixed height and width.
Note that the merge operation creates a new spike train where the distance between spikes cannot be described using Eq.~\ref{eq:dirac_comb}, but instead requires the use of Eq.~\ref{eq:spike_train}.

\subsection{Fourier Transform}
\label{subsec:fft}
In the software developed for this work, spikes are represented by ones in a tensor of zeros with the length of the tensor being as large as the simulation time multiplied by the sampling frequency (or simulation clock).

These spike tensors can be interpreted as continuous time-varying signals, in which case each individual spike cannot be described by a Dirac $\delta$ but needs to be associated to a specific width and a unitary height. This results in a sequence of rectangles ($\mathrm{rect}$) in time.

From this, the frequency composition of the spike train can be obtained by examining the spectral footprint generated by the \gls{ft}, which would be impossible when using the Dirac $\delta$ representation.

\subsection{Synthetic Dataset}
\label{subsec:synthetic-dataset}
To evaluate the performance of various algorithms at identifying different frequencies, we created a dataset of spike trains composed of two frequencies, each lasting \SI{100}{\milli\second}, as depicted in Figure~\ref{fig:accuracy_comparison_isi_fourier}(A). The first frequency $F_1$ was varied between \SI{30}{\hertz} and \SI{54}{\hertz} with a step of \SI{3}{\hertz}, while $F_2$ had the same range but a step of \SI{9}{\hertz}, resulting in a total of 27 unique combinations. Each combination was created 100 times. Each spike train was affected by a temporal shift (randomly selected but fixed for each train, e.g. affecting all spikes equally) and jitter noise. The noise was sampled for every spike from a normal distribution:
\begin{equation}
\sigma(f,k) = \mathcal{N}(0,1)\frac{\mathrm{mag_\sigma}f_N}{f}
\end{equation}
whereas the shift was computed as:
\begin{equation}
\mu(f) = \mathcal{N}(0,1)\frac{\mathrm{mag_\mu}f_N}{f}
\end{equation}
with $\mathrm{mag_\sigma}$ = \SI{100}{\milli\second} and $\mathrm{mag_\mu}$ =  \SI{1}{\milli\second}. In these equations, $f_N$ is the Nyquist sampling frequency, $f$ represents the frequency of the spike train, and k is the index of the single spike. 

The dataset was divided in training (2187 samples, 81\%), evaluation (243 samples, 9\%) and testing (270 samples, 10\%). The distinction between training and evaluation was only present during hyperparameter optimization (Sec.~\ref{subsec:nni}).

\subsection{Classifier} 
\label{subsec:classifier}
To assess the performances of the different algorithms against the syntethic dataset described above, a classifier was introduced. A linear layer of 27 output units was benchmarked with one-hot encoding (i.e, the label of the data is compared with the index of the most active neuron). The weights of the linear classifier were updated using traditional backpropagation, instructed by a cross-entropy loss~\cite{Mao_etal23}. 

Due to the one-hot encoding requiring a single output label, a strategy consisting of creating one class per frequency combination was selected, since compared to the usage of multi-label classifiers, this method offers the advantage of reducing the hyperparameters to be optimized (specifically the output activity threshold).

\subsection{Backpropagation Through Time}
In this work, specifically for the model's multifrequency experiment, we trained the weights of the spiking neural networks. To do so we used \acrfull{bptt}~\cite{Werbos1988}, an algorithm for applying backpropagation to functions that have a temporal dynamics. 

The \gls{bptt} algorithm first computes the activity of the neuron with a normal iterative process (in this work forward Euler), allowing for the simulation of the network of neurons. It then activates the backward pass, which consists of computing the error of the network through the loss function, after which the error is  propagated back to the network's elements. The algorithm saves all the time steps computed in the forward pass and propagates the gradient in the temporal domain, unfolding the network as if it was a multi-layer network (unroll). 

In this work, the \gls{bptt} is automatically setup by the autograd function of Pytorch~\cite{Paszke_Gross_ea2019}.

\subsection{Surrogate Gradient}
To train spiking neural networks we need to backpropagate the error computed with the loss function through the activity of the neurons. However, the spikes that neurons use to communicate cannot be differentiated. In this work we use the surrogate gradient proposed in~\cite{Zenke_Ganguli2018} to overcome the problem. In this method, the spike is convoluted with a kernel composed of the normalized negative part of a fast sigmoid such that the gradient is
\begin{equation}
 \nabla_{\mathrm{surrogate}} = \frac{\nabla_{\mathrm{in}}}{\left(a \cdot|\mathrm{spk}| + b\right)^2}
\end{equation}
where $a = 20$ and $b=1$, as used in~\cite{Muller-Cleve_Fra_ea2022}. Thanks to this operation, the algorithm can compute a continuous gradient in the backward pass, while simulating a normal spiking neuron in the forward pass.
\subsection{Hyperparameter Optimization}
\label{subsec:nni}
The models used in this work (explained in Section~\ref{sec:models}) have several parameters that influence the performance of the network. For example, the \gls{dsnn} model presented in Sec.~\ref{subsec:lif_network} has $\tau$s and gains that affect the frequency or the excitability of the neurons. To guarantee a fair comparison across models, we optimized the \gls{r-lif} networks using a hyperparameter optimization framework called \gls{nni}~\cite{nni2021}. This program runs the desired script several times, while pursuing the maximum accuracy. In the \gls{nni} framework, the evaluation part of the dataset was used as testing. 

For the \gls{dsnn} the \gls{nni} optimized $\tau_{\mathrm{syn}}$, $\tau_{\mathrm{neu}}$, $\mathrm{gain}_{\mathrm{syn}}$, $\mathrm{gain}_{\mathrm{neu}}$, $lr_{\mathrm{R-LIF}}$ and $lr_{\mathrm{dec}}$. The last two elements are the learning rate of the spiking neurons and the learning rate of the decoder, respectively. The algorithm was run 100 times using annealing. For the \gls{spll}, no automatic optimization has been performed due to the manual optimization already performed in the past. For the \gls{lstm}, no optimization algorithm was used.

\subsection{Potential Well Analysis}
\label{subsec:potwell}
To gain insights on the \gls{spll} behavior, we devised a model based on the concept of a potential landscape in phase space~\cite{Tiemens2023}. The potential $V(\phi)$ is defined with respect to the phase difference of the system (Sec.~\ref{subsec:spll}):
\begin{equation}
\phi = \phi_{\mathrm{IN}}-\phi_{\mathrm{CCO}}\pmod{2\pi}
\end{equation}
such that $\phi$ is positive if the \gls{cco} spikes after the input IN. One can imagine to place a ``phase particle'' on this potential landscape and observe the evolution of the system. The ``phase particle'' would move in the energy landscape, seeking the minimum of the potential energy within the cyclic phase space.

An illustrative example of such a potential landscape is presented in Figure~\ref{fig:spll}(D). To comprehend how this landscape was created, we first need to define several fictitious forces. The first force, $F_{\mathrm{freq}}$, is a constant term related to the intrinsic frequency difference between the \gls{cco} and the input IN:
\begin{equation}
F_{\mathrm{freq}} \propto f_\mathrm{IN}-f_\mathrm{cco}
\end{equation}
However, this force alone would cause $\frac{d\phi}{dt}$ to tend to infinity, necessitating the introduction of a friction term, $C_f\frac{d\phi}{dt}$, where $C_f$ is an arbitrary constant. Finally, we introduce $F_{\mathrm{TDE}}$, a force modeling the behavior of the \gls{tde}, the spikes of which accelerate the \gls{cco}. This gives us the complete equation of motion:
\begin{equation}
\label{eq:eom}
m\frac{d^2\phi}{dt^2} = F_{\mathrm{freq}} - C_f\frac{d\phi}{dt} + F_{\mathrm{TDE}}
\end{equation}
where $m$ is the ``mass'' of our phase particle, related to the dynamics of the \gls{cco} synapse. To find the potential, we integrate the conservative forces with respect to the phase $\phi$:
\begin{equation}
V(\phi) = -\int \left(F_{\mathrm{freq}} + F_{\mathrm{TDE}}\right) \ d\phi
\end{equation}
As we cannot measure these forces directly however, we substitute Equation~\ref{eq:eom} to obtain:
\begin{equation}
V(\phi)= -\int \left(m\frac{d^2\phi}{dt^2} + C_f\frac{d\phi}{dt}\right) \ d\phi
\end{equation}
Finally, Figure~\ref{fig:spll}(D) was generated by simulating the model multiple times with varying initial conditions, and determining $\frac{d\phi}{dt}$ and $\frac{d^2\phi}{dt^2}$ for each run. Subsequently, all the data points were numerically integrated together.

\printbibliography
\onecolumn
\newpage
\appendix
\renewcommand\thefigure{S\arabic{figure}}
\setcounter{figure}{0}    
\section*{Supplementary Materials}
\subsection*{Instantaneous Rate Histogram}
\label{subsec:irh}
When analyzing the information contained in spike trains, a typical approach used in literature is the \acrfull{irh}. This algorithm computes the distribution of the time between each consecutive spike in the spike train.

Using Eq.~\ref{eq:spike_train}, we can express it as follows:
\begin{equation}
 R(t) = \frac{1}{\sum_k t_{s}[k]} \sum_k {\delta(t-t_{s}[k]+\sigma[k] + \mu)} \xrightarrow{ISI} t_s+\sigma[k] + \mu \xrightarrow{HIST} \mathrm{dist}(t_s+\sigma[k] + \mu)
 \label{eq:6}
\end{equation}

Given a spike train with a specific frequency, the histogram of the reciprocal of the \gls{isi}, here defined as \gls{irh}, is likely to produce a Gaussian function around the spike train's frequency (for example in Eq.~\ref{eq:spike_train} it is gonna reveal $mean(t_s)$).

In the case of multiple frequencies, the resulting \gls{isi} creates a new $t_s$. Using a probability analysis, we can infer what the distribution of \gls{isi} looks like. Assuming two spike trains $R_1$ and $R_2$ with periods $T_2>T_1$, the likelihood of having a spike generated by $R_1$ is given by:
$$ P(T_1) = \frac{\alpha}{T_1} \left[\left(1-\frac{T_1}{T_2}\right)\delta(t-T_1) + \frac{T_1}{T_2}\mathcal{U}(0,T_1)\right]$$
The probability of having a spike generated by $R_2$ is given by:
$$ P(T_2) = \frac{\alpha}{T_2}\mathcal{U}(0,T_1)$$
where $\alpha$ is such that $\alpha/T1 + \alpha/T2 = 1$.

The resulting distribution comes out as:
\begin{equation}
 P = P(T1) + P(T2) = \frac{2T_1}{T_1 + T_2}\mathcal{U}(0,T_1) + \delta(t-T_1)
 \label{eq:spike_distr}
\end{equation}

The histogram in this case gives a blurry picture about which frequencies were initially encoded in the spike trains. The histogram has been computed using the Numpy module's histogram function.

\subsection*{Fourier Transform and \acrlong{isi}}
\label{subsec:isivsft}

Information about the internal state of the neuron is typically propagated using the reciprocal of the time difference between spikes, denoted as \gls{ir} ($\text{IR} = 1/\text{ISI}$), as explained in Sec.~\ref{subsec:spikes}. In this context, the neuron's internal state at time $t$ is characterized by the reciprocal of the time difference between $\delta_{t}$ and $\delta_{t-1}$. The mean of \gls{ir} provides the average neuron firing rate, thus informing about its the input signal.

However, when a spike train contains multiple variables to transmit, analyzing the single instantaneous rate might be insufficient. The time difference between every consecutive spike could be misleading, particularly when introducing the concept of $t_s$ in a merged spike train (as discussed in~\ref{subsec:spikes}).

\begin{figure*}[tbh]
 \centering
 \includegraphics[width=1\linewidth]{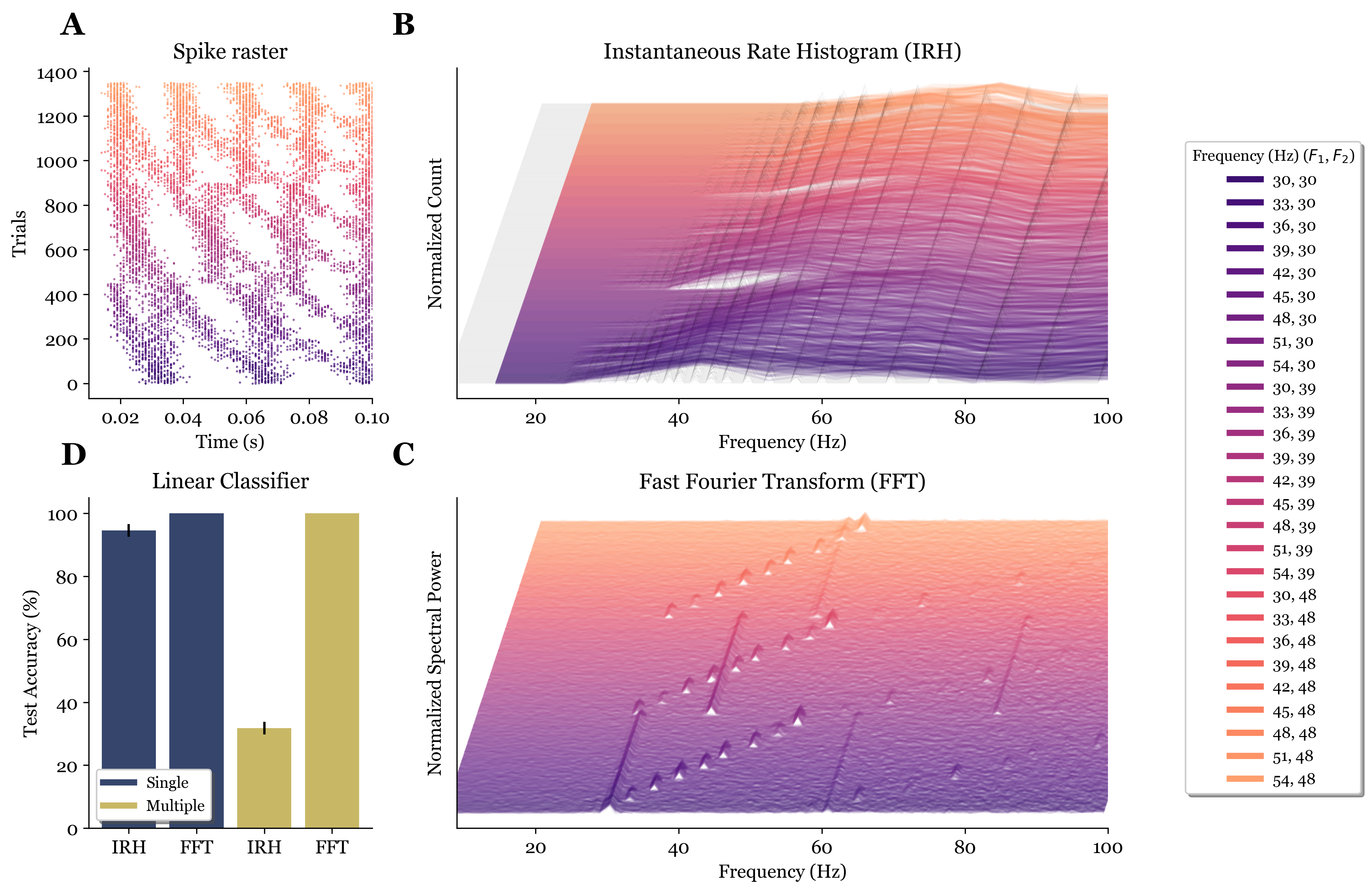}
 \caption{In panel \textbf{A} the raster plot of the dataset introduced in Sec.~\ref{subsec:synthetic-dataset} is illustrated. In panel \textbf{B} the \gls{irh} for the dataset in A is shown with a bin number of 10. In panel \textbf{C} an \gls{fft} of the spike trains is computed for every trail with a sampling rate of \SI{1}{kHz}. Finally in panel \textbf{D}, the results of a linear classifier predicting the frequency composition in a single frequency and a double frequency case for 15 different seeds. Both the \gls{irh} and \gls{fft} have a dimensionality of 200.}
 \label{fig:accuracy_comparison_isi_fourier}
\end{figure*}

To investigate this, we conducted a benchmark comparing two different spike analysis methods: utilizing the \gls{irh} of the spike train and employing the \gls{fft} of the spike train (as explained in Sec.~\ref{subsec:fft}). The dataset used for this analysis comprises two frequencies with added jitter noise, representing two internal variables ($F_1$ and $F_2$), elaborated further in Sec.~\ref{subsec:synthetic-dataset} and depicted in Figure~\ref{fig:accuracy_comparison_isi_fourier}(A).

For the \gls{irh}, we computed the time difference between each consecutive spike and then applied a histogram function. The results, depicted in Figure~\ref{fig:accuracy_comparison_isi_fourier}(B), exhibit an interplay between the two frequencies. The plot reflects the behavior outlined in Sec.~\ref{subsec:spikes}, where for each frequency combination, the lowest frequency determines the minimum x-value of the histogram. Meanwhile, the upper part of the histogram is evenly filled with events due to the random shift between the two frequencies. It's worth noting that introducing random jitter to both frequencies results in a Gaussian spread around the peak predicted by the equations. This outcome underscores the challenge of identifying the frequencies composing a given histogram, especially in the presence of random noise. Note that the black lines depict the histogram if 200 bins are used, while the colored lines depict the case for 10 bins. The 10 bins case has been used to improve readibility while the 200 bins case has been used to test against the \gls{fft} (explained further).

In contrast, Figure~\ref{fig:accuracy_comparison_isi_fourier}(C) portrays the same dataset analyzed using the \gls{fft}. Here, the spikes are considered as ones and the absence of spikes as zeros. The Fourier transform is able to consistently identify the original frequencies and distinguish the different contributions of the two internal variables. Harmonics of the frequencies are also visible.

To quantitatively test this hypothesis, we fed the outputs of \gls{irh} and \gls{fft} to a linear classifier for both the described dataset and a dataset with only one frequency (27 instances from \SI{30}{\hertz} to \SI{54}{\hertz}). This was repeated by varying the random seed 15 times. Both the \gls{irh} and \gls{fft} had a dimensionality of 200. The classifier was trained with these data for 100 epochs as explained in Sec.~\ref{subsec:classifier}. As illustrated in Figure~\ref{fig:accuracy_comparison_isi_fourier}D, for the single frequency example, both the \gls{fft} and \gls{irh} are able to discern the frequency in a comparable manner. This is because the interspike interval reflects the input frequency. However, when adding the second frequency to the input, the \gls{fft} algorithm can more reliably discern the different frequencies present in the dataset, while the \gls{irh} struggles. This can be attributed to the fact that merging several spike trains makes the interspike interval unreliable in determining the contained frequencies.

The primary contribution of this experiment is demonstrating that multiple variables can be encoded in a single spike train, with each variable's value represented by a spiking frequency. We have demonstrated that standard procedures like \gls{irh} may not adequately extract or visualize this information, necessitating more complex analyses.
\end{document}